\documentclass[reprint,showpacs,footinbib,citeautoscript,floatfix,nofootinbib]{revtex4-1}

\usepackage{graphicx}
\usepackage{amsmath}
\usepackage{amssymb}
\usepackage{mathrsfs}
\usepackage{bm}
\usepackage{array}
\usepackage{natbib}
\usepackage{soul}
\usepackage{natbib}

\newcolumntype {s}[1]{@{\hspace{#1}}} 
\usepackage{color}
\usepackage[usenames,dvipsnames]{xcolor}
\usepackage{wrapfig}
\usepackage[normalem]{ulem}

\begin{document}

\title{Spin polarisation and non-isotropic effective mass in the conduction band of GdN}

\author{W.~F.~Holmes-Hewett$^{1,3,*}$ ,E.~X.~M.~Trewick$^{2,3,}$, H.~J.~Trodahl$^{2}$, R.~G.~Buckley$^{1,3}$ and B.~J.~Ruck$^{2,3}$}

\affiliation{$^{1}$Robinson Research Institute, Victoria University of Wellington, P.O. Box 33436, Petone 5046, New Zealand}

\affiliation{$^2$School of Chemical and Physical Sciences, Victoria University of Wellington, P.O. Box 600, Wellington 6140, New Zealand}

\affiliation{$^3$MacDiarmid Institute for Advanced Materials and Nanotechnology, P.O. Box 600, Wellington 6140, New Zealand}

\affiliation{*Corresponding author: W.~F.~Holmes-Hewett, William.Holmes-Hewett@vuw.ac.nz}

\date{\today}


\begin{abstract}

GdN is a ferromagnetic semiconductor which has seen increasing interest in the preceding decades particularly in the areas of spin- and superconducting- based electronics. Here we report a detailed computational study and optical spectroscopy study of the electronic structure of stoichiometric and nitrogen vacancy doped GdN. Based on our calculations we provide the effective mass tensor for undoped GdN, and some indicative values for electron doped GdN. Such a property is valuable as it can directly affect device design, and be directly measured experimentally to validate the existing computation results. 

\end{abstract}

\maketitle

\section{Introduction}

The increasing global demand for computational efficiency~\cite{Andrae2015} has pushed the development of conventional CMOS based logic and memory beyond the imagination of the materials scientists and physicists who pioneered the first generation of transistor based devices decades ago. With the limits of the present family of conventional-materials based devices (perpetually) on the horizon there has been an ongoing effort to design more exotic logic and memory systems, systems which can surpass the conventional technologies leading to orders of magnitude increases in operating efficiency and speed~\cite{IRDS2018-BCMOS,IRDS2018-MM}.

Among the most attractive of these technologies are spintronics~\cite{Hirohata2020,Bhatti2017} and superconducting electronics~\cite{Soloviev2017,Likharev1991}. Both of these operate at cryogenic temperatures, temperatures which limit the use of conventional materials. As such, a search for materials with the functionalities required for these technologies and the ability to operate at the required temperatures is an active area of materials science~\cite{Li2016,Alam2023}. In this context the rare-earth nitride (LN - L a lanthanide) series of ferromagnetic semiconductors~\cite{Natali2010} has seen a burst of interest in recent years. First studied in the 1950s~\cite{Eick1956} the LN were found to be mostly ferromagnetic with Curie temperatures on the order of 50~K~\cite{Hulliger1978}. The simple rock-salt structure and limited quenching of the orbital angular momentum results in rich and varied magnetic behaviours across the series, which is driven by the 4\textit{f} occupation ranging from 0 in LaN to 14 in LuN. The series was initially thought to be half-metallic, similar to the other rare-earth pnicitdes~\cite{Hulliger1979}. However, improved experimental procedures allowing the growth of high quality thin films, along with improved computational methods, has in recent decades revealed the semiconducting nature of at least several members of the series~\cite{Granville2006,Aerts2004,Anton2016b,Holmes-Hewett2020,holmes-hewett2023,Devese2022}. 

\begin{figure}
\centering
\includegraphics[width=0.8\linewidth]{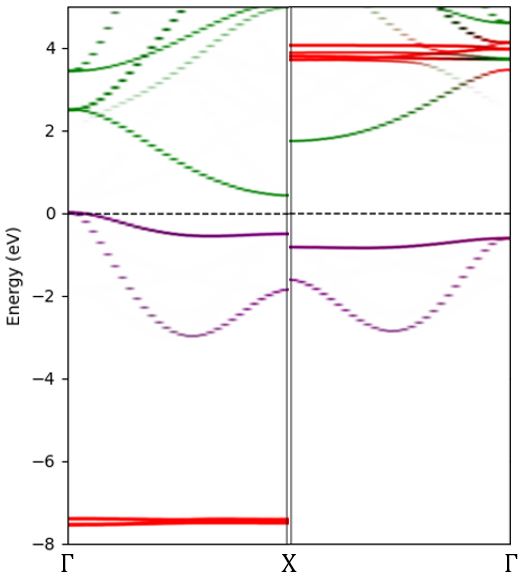}
\caption{Calculated band structure of stoichiometric GdN (left - majority spin, right - minority spin) along the line $\Gamma$-X. The wavefunctions are projected onto the atomic orbitals, Gd~5\textit{d} - green, Gd~4\textit{f} - red N~2\textit{p} - purple.}
\label{Bands1}
\end{figure}

GdN is the most studied of the rare-earth nitrites, with a Curie temperature reported between 50~K and 70~K~\cite{Ludbrook2009}. The half filled 4\textit{f} shell yields a strong magnetic moment of 7~$\mu_B$ per Gd ion, or 32~$\mu_B$ per nm$^3$. GdN is insulating in the stoichiometric state with a valence band formed from N~2\textit{p} states and a conduction band minimum formed from Gd~5\textit{d} states. The stoichiometric band structure calculated in the present study is shown in Figure~\ref{Bands1}. The valence band maximum is at $\Gamma$ and there is a $\sim$~0.9~eV optical band gap between the majority spin states at the X point~\cite{Trodahl2007}, to which we have tuned our calculation (see section~\ref{Methods}). The minority spin gap is significantly larger in the calculation ($\sim$~2.5~eV) between the minority spin N~2\textit{p} and Gd~5\textit{d} states at X, a result of a large spin-splitting in the conduction ($1.3$~eV) and the valence ($0.2$~eV) bands both resulting from exchange with the Gd~4\textit{f} states~\cite{Larson2006}. Previous computational studies~\cite{Larson2007,Preston2010a}, also tuned to the optical gap in the ferromagnetic phase, find a significantly smaller minority spin gap of $\sim~1.7$~eV. 

The onset of optical absorption in the ferromagnetic phase represents the majority spin gap, as both the conduction band minimum and valence band maximum are majority spin. There is largely agreement in the existing experimental studies regarding the 0.9~eV majority spin gap~\cite{Trodahl2007,Vilela2024,Azeem2016,Yoshitomi2011}. There is however much less agreement regarding the minority spin gap and associated spin-splitting, the latter of which is reported variously as 0.16~eV, 0.4~eV and 0.6~eV respectively in Refs~~\cite{Vidyasagar2012,Azeem2016} and \cite{Vilela2024}. In the context of majority-minority spin gaps the paramagnetic gap can be estimated as an average of the two. The present calculation shown in Figure~\ref{Bands1} yields 1.75~eV, somewhat above the experimental paramagnetic gap of 1.5~eV. 

As with many of the other rare earth nitrides GdN can be doped with electrons though the inclusion of nitrogen vacancies~\cite{Maity2018,Holmes-Hewett2020,Devese2022}. These have a low activation energy, dependent on growth conditions~\cite{Punya2011}, so can be readily formed in thin films to the order of a few \% by either reducing the N$_2$ partial pressure during deposition, or controlling the substrate temperature. Experimental results on intentionally doped GdN films support the simple argument of the nitrogen vacancy donors raising the Fermi level into the (initially) majority spin only Gd~5\textit{d} bands~\cite{Trodahl2017}.

\begin{figure*}
\centering
\includegraphics[width=\linewidth]{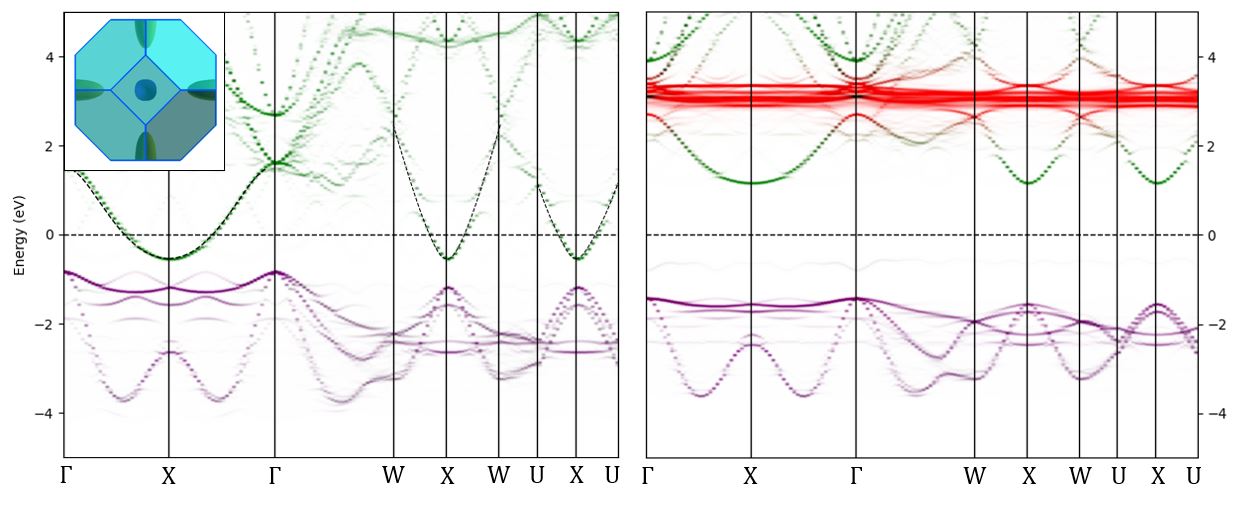}
\caption{Calculated band structure for electron doped GdN$_{1-\delta}$ highlighting the dispersion around the conduction band minimum at X: left - majority spin, right - minority spin. The wave functions are projected onto the atomic orbitals, Gd~5\textit{d} - green, Gd~4\textit{f} - red N~2\textit{p} - purple. The Gd~5\textit{d} states from the stochometric crystal are shown in the left hand panel in black dashed lines. The inset shows the Fermi surface highlighted on the first Brillouin zone.}
\label{Bands2}
\end{figure*}

The rare confluence of a combine semiconducting-ferromagnetic ground state has seen GdN included in many prototype cryogenic devices. The material is particularly attractive in the area of superconducting spintronics. The inclusion of ferromagnetic materials as the weak link in a Josephson junction structure adds functionality based on the hysteric nature of the magnetisation~\cite{Birge2024}, however the low resistance of the ferromagnetic-metals commonly used results in switching speeds orders of magnitude lower than in conventional superconducting logic, rendering conventional ferromagnetic-metal based Josephson junctions incompatible~\cite{Vernik2013}. In this context GdN has been used as the ferromagnetic insulator of choice in the superconducting spintronics community, with various devices already demonstrated including superconducting- spin-filters~\cite{Senapati2011,Massarotti2015,Ahmad2020}, diodes~\cite{Sharma2023}, primitive memory devices~\cite{Cascales2019,pot2023}, and even the recent proposal of a GdN based \textit{ferro-transmon} qubit~\cite{Ahmad2022b}. 

Given the roughly a decade of intense interest in the application of GdN in such devices it is surprising that the most recent computational studies regarding the electronic structure of GdN are themselves over a decade old~\cite{Larson2007,Punya2011,Mitra2008}. In the present manuscript we attempt to fill this void by undertaking a computational study of GdN, informed by experiment, where we aim to provide details regarding electron transport in GdN and realistic material parameters which will aid experimentalists and engineers in device design.

\section{Methods}
\label{Methods}




\subsection{Computational}

Density functional theory (DFT) based calculations were undertaken using Quantum Espresso~\cite{QE,Cococcioni2005} and rare earth pseudo-potentials developed using the rare-earth nitrides~\cite{Topsakal2014}. Self-consistent calculations on the primitive cell were completed using a $k$-mesh with $10\times10\times10$ divisions, while super-cell calculations were on a $4\times4\times4$ division \textit{k}-mesh. The wave function and charge density cut-off energies were 50 Ry and 200 Ry respectively for all calculations. Following our DFT calculations the output from Quantum Espresso was used to generate maximally localised Wannier functions using Wannier~90~\cite{Mostofi2014,Marzari1997,Souza2001}. The resulting Wannier functions were then used to calculate the electronic properties on denser \textit{k}~meshes, a $25\times25\times25$ grid for calculation of the density of states and Fermi level) and a $125\times125\times125$ grid for the optical conductivity. The effective mass was determined using conventional DFT based on 400~\textit{k} along each high symmetry direction.

The 4$f$ electrons of the \textit{Ln}N series are strongly correlated and thus require careful treatment beyond the traditional DFT methods~\cite{Larson2006,Larson2007}. In the basic DFT the 4$f$ states are found at or near the Fermi level for most of the stoichiometric rare earth nitrides, when in reality this is not the case. The strongly correlated nature of these electrons pushes the filled states below and unfilled states above the Fermi level. This physics can be approximated using the DFT+$U$ method where the behaviour of the correlated orbitals is determined by an adjustable parameter $U$. In the present study two $U$ parameters are used, as described first in reference~\cite{Larson2006}. One to account for the strongly correlated 4$f$ states ($U_f$), and a second applied to the 5$d$ states ($U_d$) to adjust to the experimental optical band gap in the ferromagnetic phase. The process is discussed further in references~\cite{Holmes-Hewett2021} and~\cite{holmes-hewett2023}.

\subsection{film growth and optical spectroscopy}

GdN thin films on the order of 100~nm were prepared in a Thermionics ultrahigh vacuum chamber with a base pressure of $10^{-9}$~mbar. Gd metal was evaporated by an electron gun in a N$_2$ partial pressure of $1\times 10^{-4}$~mbar at a rate of $\sim1$~\AA~s$^{-1}$. Varying the N$_2$ growth pressure acts to control the concentration of nitrogen vacancies (V$_N$) in the films. The films were capped with $\sim50$~nm insulating AlN to minimise deterioration when exposed atmospheric water vapour and oxygen. The films were simultaneously deposited at ambient temperature on Si and Al$_2$O$_3$ substrates to suit the required measurements.

Samples for electrical transport measurements were deposited on $10~\times~10$~mm$^2$ Al$_2$O$_3$ substrates predeposited with Cr/Au contacts in the van der Pauw configuration. Electrical measurements were conducted in a Janis closed-cycle helium cryostat and a Quantum Design Physical Properties System at temperatures from 300 to 4~K. Reflection and transmission measurements were performed with a Bruker Vertex 80v Fourier transform spectrometer in the 0.012-3~eV spectral range using films deposited on both Si and Al$_2$O$_3$ substrates. The reflectivity was measured relative to a 200~nm Al film and then corrected for the finite reflectivity of Al using data from Ehrenreich \cite{Ehrenreich1963}. The reflectivity and transmission spectra were simultaneously modelled with refFIT software \cite{RefFit} using a Kramers-Kronig consistent consistent sum of Drude-Lorentzians to calculate the energy-dependent dielectric function from which the presented optical conductivity data was generated. 

\section{Results}

\subsection{Density function theory}
\label{DFT-section}
As is commonly reported, GdN is grown with varying nitrogen vacancy doping to control the carrier concentration. In the doped crystal the Gd ions in the lattice retain their 3+ oxidation state, so to maintain charge neutrality the three electrons formerly on the nitrogen ion must be accounted for. Two electrons occupy defect states which localise to the vacancy site, while the third electron finds a higher energy defect state, drawing the Fermi level up into the extended state Gd~5\textit{d} conduction band. This one mobile electron per nitrogen vacancy is responsible for electron transport. The Fermi surface is then a prolate spheroid with its centre at the X point, major axis pointing along X$\Gamma$ and minor axes in the XW and XU plane. The first Brillouin zone with the Fermi surface highlighted is shown in the inset to Figure~\ref{Bands2}. It is then the dispersion along these three high symmetry directions which will dictate the transport properties of the material.

To represent the above description Figure~\ref{Bands2} shows the calculated band structure of nitrogen vacancy doped GdN along these high symmetry directions. It is clear that the curvature of the bands contrasts significantly, resulting in a differing effective mass, and ultimately transport properties along different high symmetry directions. The figure shows the calculated wave-functions projected onto the atomic orbitals (colours noted in caption), in addition we have calculated the dispersion for the stoichiometric Gd~5\textit{d} conduction band along the same high symmetry directions, shown for the majority spin as black dashed lines in the left hand panel of Figure~\ref{Bands2}. It is significant that the stoichiometric and nitrogen vacancy doped conduction bands are essentially superimposed on one another, fitting the picture that the electrons donated by nitrogen vacancy doping are simply lifting the Fermi level into the intrinsic conduction band. Although the shape of the bands is retained between the stoichiometric and doped calculations the gap between the Gd~5\textit{d} and N~2\textit{p} states at the X point has reduced from $\sim$~0.9~eV in the stoichiometric case to $\sim$~0.6~eV in the doped case, the renormalzation likely a result of screening from the delocalised electrons. The onset of optical absorption is still near $\sim$~0.9~eV, however now between $\Gamma$ and X, and expected to be much weaker as the initial states are the occupied defect states, here with a concentration of only 1/27 to that of the N~2\textit{p}. The \textit{intrinsic} majority spin optical band gap (i.e. not involving defect states) increases to $\sim$~1.2~eV ($\sim$~0.9~eV in the stoichiometric crystal) in a Moss-Berstein like shift. 

The minority spin states are shown in the right hand panel of Figure~\ref{Bands2}. The Gd~5\textit{d} states are again very similar to the stoichiometric crystal. However, the presence of the occupied defect states in the gap, below the Fermi level, effectively reduces minority spin optical gap from 2.5~eV in the stoichiometric to 1.8~eV in the doped crystal. The average absorption onset of the majority and minority spin gaps in the doped crystal (both with initial impurity states and Gd~5\textit{d} final states) is now $\sim$~1.35~eV (1.75~eV in the stoichiometric crystal). This is now less than is commonly reported for the paramagnetic band gap. 

Using the present results from the doped crystal the spin splitting can be reported in various ways. The most appropriate for electrical transport (i.e. as in a spin filter) is the minimum energy difference between available (hole or electron) majority and minority spin states. As Fermi level lies in the majority spin Gd~5\textit{d} band, for electron carriers this is the energy difference between the Fermi level and the bottom of the minority spin conduction band, roughly 1.1~eV, while for hole carriers this is the energy difference between the Fermi level and the top of the minority spin valance band, roughly 0.6~eV. For optical studies the most relevant spin splitting is the difference between the majority spin and minority spin optical gaps, which  now $\sim$0.9~eV. This is closer to the experimental reports, yet still beyond the maximum 0.6~eV reported in reference~\cite{Vilela2024}.


\subsection{Optical spectroscopy}

\begin{figure}
\centering
\includegraphics[width=\linewidth]{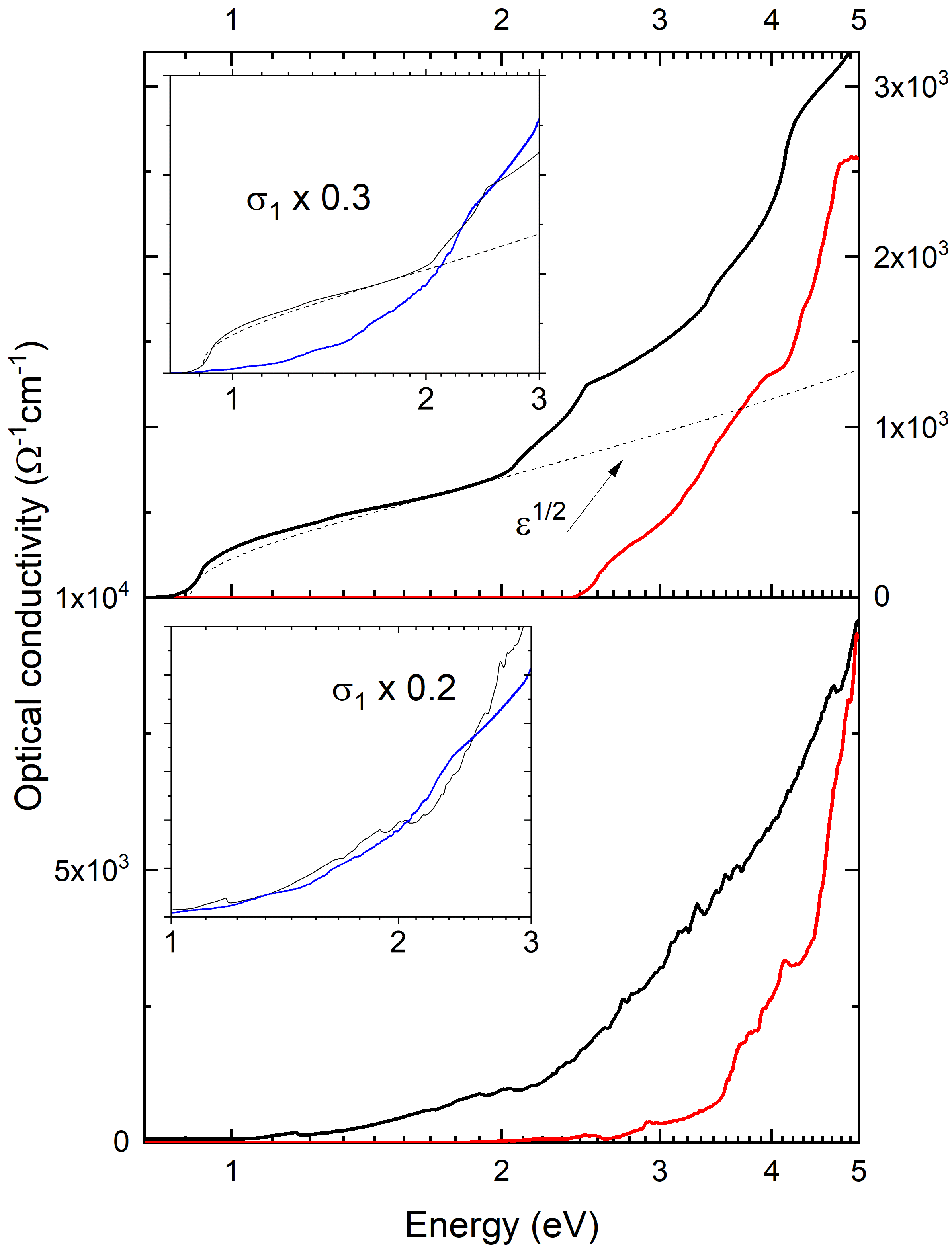}
\caption{Calculated optical conductivity for pristine (top panel) and nitrogen vacency doped (bottom panel) GdN. Both majority (black) and minority (red) spin contributions are shown. The insets show the majority spin contribution and in addition experimental data collected at 8~K.}
\label{Optical}
\end{figure}


As an initial comparison of our computational results to experiment we have calculated the optical conductivity for both the pristine and doped cells discussed above. These are displayed in Figure~\ref{Optical}. The top panel of Figure~\ref{Optical} shows the calculated optical conductivity for stoichiometric GdN. The onset of optical absorption is in the majority spin states at $\sim$~0.9~eV and initially shows an increase with a near $A+(\epsilon-E_g)^{1/2}$ form, which is expected from the Van Hove singularity at the conduction band minimum. Above 2~eV the gradient increases relating to the flattening of both the 5\textit{d} and the N~2\textit{p} bands near +2~eV and -2~eV respectively. At roughly 2.5~eV there is a similar onset of absorption in the minority spin states. The inset show the spectra between 0.8~eV and 3~eV now with experimental data (blue) measured at 8~K. Although the onset of absorption is a roughly the same energy in both spectra the form is very different, with the experiment not taking the typical shape expected for the onset of optical transitions in a parabolic band.

The lower panel of Figure~\ref{Optical} shows the calculated optical conductivity for the nitrogen vacancy doped super-cell. Here, as discussed in section~\ref{DFT-section}, the absorption onset is at roughly the same energy (0.9~eV) as the pristine cell. For the doped material the form of the optical conductivity differs greatly from the simple $\epsilon^{1/2}$ shape as the Fermi level now lies within the conduction band. The inset again shows the comparison between the calculation and experiment with a clear qualitative similarity between the two (the calculation has been scaled by a factor of 0.2 to best match the experiment). The film used for the measurement was moderately conductive with a resistivity of 0.1~$\Omega$cm at room temperature, increasing to 10~$\Omega$cm at 4~K. Given this resistivity, and based on Hall measurements on other films~\cite{Lee2015}, one could assume a carrier concentration on the order 10~$^{17}$cm$^{-3}$ at 4~K, however the striking resemblance between the optical spectra for this film and the super-cell calculation (carrier concentration $\sim 1~\times 10^{21}$cm$^{-3}$) may rather imply that in this is incorrect. This film and possibly other nominally \textit{undoped} rare-earth nitrides may harbour a significant population of localised carriers. These do not contribute to electron transport in the usual extended state manner; however, they do occupy states at the bottom of the conduction band, affecting the optical spectra and other related electrical properties. Similar optical spectra have been seen in other rare-earth nitrides~\cite{Holmes-Hewett2019,Holmes-Hewett2020,holmes-hewett2023} and along with electrical transport measurements were discussed in the context of localised hopping type conductivity and a mobility edge scenario.

\subsection{Effective mass}

\begin{figure}
\centering
\includegraphics[width=\linewidth]{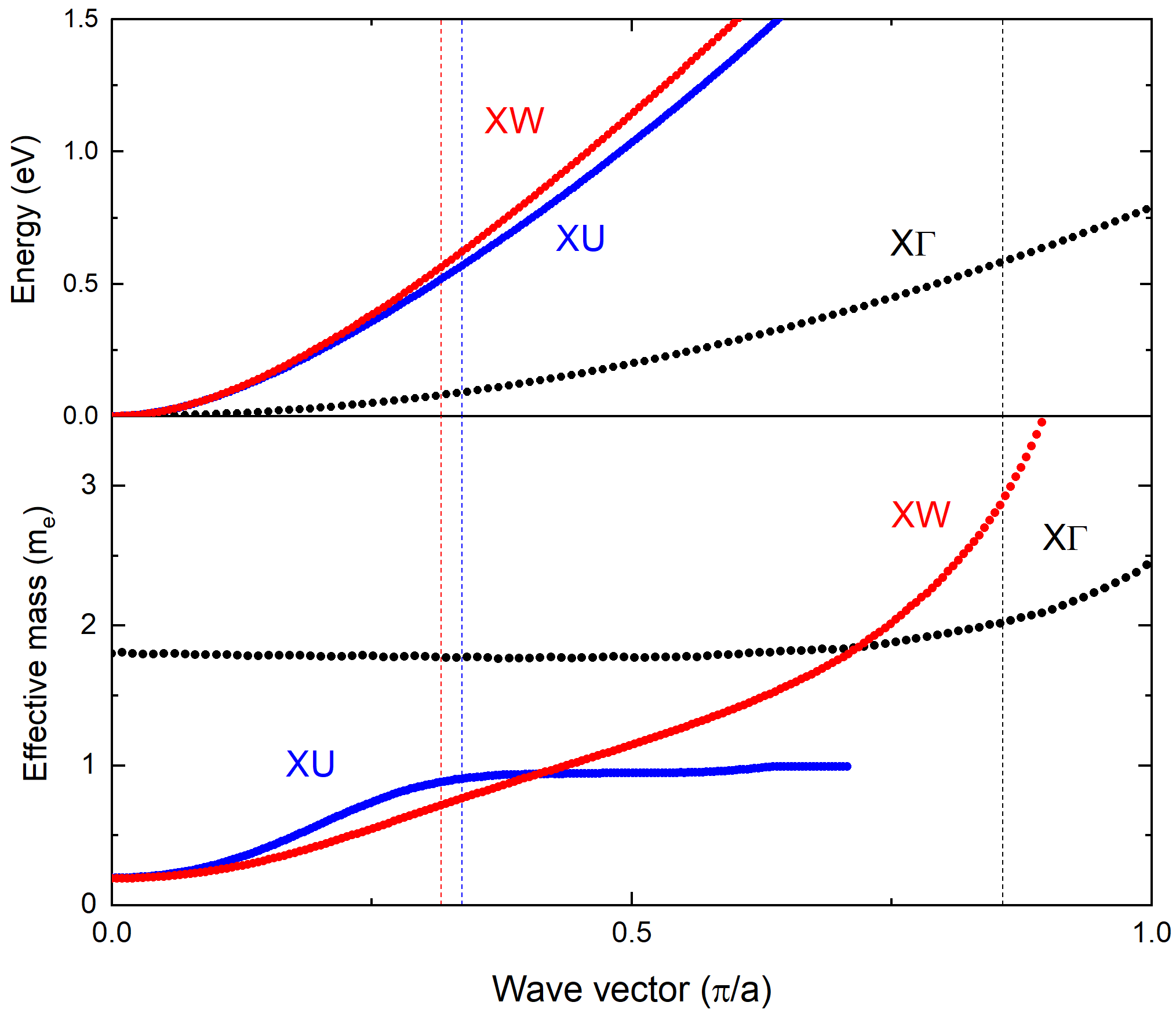}
\caption{Calculated dispersion (top) and effective mass (bottom) for GdN as a function of wave vector along various high symmetry lines with $k=$X$=0$. The Fermi wave vector for each dispersion, with reference to Figure~\ref{Bands2}, is indicated with the dashed vertical lines.}
\label{Mstar}
\end{figure}

Although the details of the spin-splitting and related exchange energy remain to be determined the present data, calculations and existing experimental reports~\cite{Trodahl2017} seem to agree that when GdN is doped with electrons via nitrogen vacancies the Fermi level is raised into the Gd~5\textit{d} band, which is majority spin only at dilute to moderate doping levels. Motivated by these observations, and the similarity between the stoichiometric and nitrogen vacancy doped Gd~5\textit{d} bands we have used the former to calculate the effective mass along the X$\Gamma$, XW and XU directions, which may be of use in the context of device design. The results are shown in Figure~\ref{Mstar} with the dispersion along the X$\Gamma$, XW and XU directions in the top panel, and the effective mass in the bottom panel, both as a function of wave-vector measured relative to the X point at $k=0$. 

At the X point the first Brillouin zone of the FCC lattice has four-fold symmetry (see inset of Figure~\ref{Bands2}), which dictates that along any two directions in the plane of the X-face the effective mass is degenerate at $k=$X. Without loss of generality we choose directions along XU and XW to form our basis, the final perpendicular direction is then along X$\Gamma$ and will have a distinct effective mass at the X point. The $k=$X effective mass tensor is diagonal in this coordinate system and given by

$$\frac{m^*}{m_e}= \begin{pmatrix}
0.18 & 0 & 0\\
0 & 0.18 & 0\\
0 & 0 & 1.8
\end{pmatrix}.$$


The enhanced curvature along the XU and XW directions results in an order of magnitude effective mass difference between the longitudinal and transverse directions of the Fermi surface. Using these values we can estimate the conductivity effective mass as $m^*_c=0.25~m_e$, which is dominated by the low mass components. Given the relative orientation between the equivalent X points in the first Brillouin zone an electrical field along any direction will result in contributions from both the transverse the longitudinal directions. The order of magnitude lower mass in the transverse direction will result in this dominating in any electrical transport measurement. 

The effective mass precisely at the X point is useful in the context of tunnelling experiments where the GdN is likely intended to be insulating and close to stoichiometric, in addition the geometry of these experiments is well controlled and for example the common case of NbN/GdN/NbN multi-layers~\cite{Senapati2011} results in transport through the GdN along a 001 direction. As discussed above the degeneracy of the X point results in contributions from both the transverse and longitudinal directions; at the X point the combination of these contributions can in principle be easily treated.

As one dopes the crystal with electrons the Fermi surface expands from the X point and the effective mass is a function of wavevector as shown in Figure~\ref{Mstar}. Along the X$\Gamma$ direction the effective mass is relatively stable with a slight drop before an increase near $k=\pi/a$. As one continues along this direction there is ultimately a sign inversion near the first Brillouin zone boundary as the curvature of the band itself changes sign.

Due to the rapidly changing mass along the two transverse directions it may be difficult to predict the transport properties of even moderately electron doped GdN. The super-cell used in the calculation of Figure~\ref{Bands2} was constructed from $3\times3\times3$ primitive unit cells with a N atom removed. The single mobile electron in a volume of $\sim915\mathrm{\mathring{A}}^3$ results in a carrier concentration of $\sim$1$\times10^{21}$cm$^{-3}$ and is representative of the heavily doped GdN from experimental reports~\cite{Maity2018}. The presence of this electron raises the Fermi level roughly 0.6~eV into the conduction band, the resulting Fermi wave vectors are indicated on Figure~\ref{Mstar}. Far from the X point the effective mass tensor will no longer take the simple diagonal form presented above, but even so the wave vector dependent effective mass shown in Figure~\ref{Mstar} gives a qualitative indication of the evolution of the effective mass with electron doping in GdN.

\section{Conclusions}

We have undertaken a computational study of the band structure and defect states in stoichiometric and nitrogen vacancy doped GdN. Our most direct result is the effective mass tensor at the X point, and calculation of the effective mass in the high symmetry directions moving away from the centre of this pocket. We have provided updated band structure calculations of GdN and compared these to conventional optical spectroscopy.  The comparison between conventional spectroscopy and the calculation indicates that even nominally dilutly doped GdN has evidence of a large population of localised carriers in occupying states at the conduction band minimum.

\section{Data Availability}

The data used during this study are available from the corresponding author upon reasonable request.

\section{ACKNOWLEDGMENTS}

This research was supported by Quantum Technologies Aotearoa, a research programme of Te Whai Ao – the Dodd Walls Centre, funded by the New Zealand Ministry of Business Innovation and Employment through International Science Partnerships, contract number UOO2347. The computations were performed on the R\={a}poi high performance computing facility of Victoria University of Wellington.

\bibliography{master.bib}

\begin{thebibliography}{52}%
\makeatletter
\providecommand \@ifxundefined [1]{%
 \@ifx{#1\undefined}
}%
\providecommand \@ifnum [1]{%
 \ifnum #1\expandafter \@firstoftwo
 \else \expandafter \@secondoftwo
 \fi
}%
\providecommand \@ifx [1]{%
 \ifx #1\expandafter \@firstoftwo
 \else \expandafter \@secondoftwo
 \fi
}%
\providecommand \natexlab [1]{#1}%
\providecommand \enquote  [1]{``#1''}%
\providecommand \bibnamefont  [1]{#1}%
\providecommand \bibfnamefont [1]{#1}%
\providecommand \citenamefont [1]{#1}%
\providecommand \href@noop [0]{\@secondoftwo}%
\providecommand \href [0]{\begingroup \@sanitize@url \@href}%
\providecommand \@href[1]{\@@startlink{#1}\@@href}%
\providecommand \@@href[1]{\endgroup#1\@@endlink}%
\providecommand \@sanitize@url [0]{\catcode `\\12\catcode `\$12\catcode `\&12\catcode `\#12\catcode `\^12\catcode `\_12\catcode `\%12\relax}%
\providecommand \@@startlink[1]{}%
\providecommand \@@endlink[0]{}%
\providecommand \url  [0]{\begingroup\@sanitize@url \@url }%
\providecommand \@url [1]{\endgroup\@href {#1}{\urlprefix }}%
\providecommand \urlprefix  [0]{URL }%
\providecommand \Eprint [0]{\href }%
\providecommand \doibase [0]{http://dx.doi.org/}%
\providecommand \selectlanguage [0]{\@gobble}%
\providecommand \bibinfo  [0]{\@secondoftwo}%
\providecommand \bibfield  [0]{\@secondoftwo}%
\providecommand \translation [1]{[#1]}%
\providecommand \BibitemOpen [0]{}%
\providecommand \bibitemStop [0]{}%
\providecommand \bibitemNoStop [0]{.\EOS\space}%
\providecommand \EOS [0]{\spacefactor3000\relax}%
\providecommand \BibitemShut  [1]{\csname bibitem#1\endcsname}%
\let\auto@bib@innerbib\@empty
\bibitem [{\citenamefont {Andrae}(2015)}]{Andrae2015}%
  \BibitemOpen
  \bibfield  {author} {\bibinfo {author} {\bibfnamefont {T.}~\bibnamefont {Andrae}, \bibfnamefont {A.S.G.;~Edler}},\ }\href@noop {} {\bibfield  {journal} {\bibinfo  {journal} {Challenges}\ }\textbf {\bibinfo {volume} {6}},\ \bibinfo {pages} {117} (\bibinfo {year} {2015})}\BibitemShut {NoStop}%
\bibitem [{IRD(2018{\natexlab{a}})}]{IRDS2018-BCMOS}%
  \BibitemOpen
  \href@noop {} {\emph {\bibinfo {title} {International roadmap for devices and systems - Beyond CMOS}}}\ (\bibinfo {year} {2018})\BibitemShut {NoStop}%
\bibitem [{IRD(2018{\natexlab{b}})}]{IRDS2018-MM}%
  \BibitemOpen
  \href@noop {} {\emph {\bibinfo {title} {International roadmap for devices and systems - More Moore}}}\ (\bibinfo {year} {2018})\BibitemShut {NoStop}%
\bibitem [{\citenamefont {Hirohata}\ \emph {et~al.}(2020)\citenamefont {Hirohata}, \citenamefont {Yamada}, \citenamefont {Nakatani}, \citenamefont {Prejbeanu}, \citenamefont {Diény}, \citenamefont {Pirro},\ and\ \citenamefont {Hillebrands}}]{Hirohata2020}%
  \BibitemOpen
  \bibfield  {author} {\bibinfo {author} {\bibfnamefont {A.}~\bibnamefont {Hirohata}}, \bibinfo {author} {\bibfnamefont {K.}~\bibnamefont {Yamada}}, \bibinfo {author} {\bibfnamefont {Y.}~\bibnamefont {Nakatani}}, \bibinfo {author} {\bibfnamefont {I.-L.}\ \bibnamefont {Prejbeanu}}, \bibinfo {author} {\bibfnamefont {B.}~\bibnamefont {Diény}}, \bibinfo {author} {\bibfnamefont {P.}~\bibnamefont {Pirro}}, \ and\ \bibinfo {author} {\bibfnamefont {B.}~\bibnamefont {Hillebrands}},\ }\href {\doibase https://doi.org/10.1016/j.jmmm.2020.166711} {\bibfield  {journal} {\bibinfo  {journal} {Journal of Magnetism and Magnetic Materials}\ }\textbf {\bibinfo {volume} {509}},\ \bibinfo {pages} {166711} (\bibinfo {year} {2020})}\BibitemShut {NoStop}%
\bibitem [{\citenamefont {Bhatti}\ \emph {et~al.}(2017)\citenamefont {Bhatti}, \citenamefont {Sbiaa}, \citenamefont {Hirohata}, \citenamefont {Ohno}, \citenamefont {Fukami},\ and\ \citenamefont {Piramanayagam}}]{Bhatti2017}%
  \BibitemOpen
  \bibfield  {author} {\bibinfo {author} {\bibfnamefont {S.}~\bibnamefont {Bhatti}}, \bibinfo {author} {\bibfnamefont {R.}~\bibnamefont {Sbiaa}}, \bibinfo {author} {\bibfnamefont {A.}~\bibnamefont {Hirohata}}, \bibinfo {author} {\bibfnamefont {H.}~\bibnamefont {Ohno}}, \bibinfo {author} {\bibfnamefont {S.}~\bibnamefont {Fukami}}, \ and\ \bibinfo {author} {\bibfnamefont {S.}~\bibnamefont {Piramanayagam}},\ }\href {\doibase https://doi.org/10.1016/j.mattod.2017.07.007} {\bibfield  {journal} {\bibinfo  {journal} {Materials Today}\ }\textbf {\bibinfo {volume} {20}},\ \bibinfo {pages} {530} (\bibinfo {year} {2017})}\BibitemShut {NoStop}%
\bibitem [{\citenamefont {Soloviev}\ \emph {et~al.}(2017)\citenamefont {Soloviev}, \citenamefont {Klenov}, \citenamefont {Bakurskiy}, \citenamefont {Kupriyanov}, \citenamefont {Gudkov},\ and\ \citenamefont {Sidorenko}}]{Soloviev2017}%
  \BibitemOpen
  \bibfield  {author} {\bibinfo {author} {\bibfnamefont {I.~I.}\ \bibnamefont {Soloviev}}, \bibinfo {author} {\bibfnamefont {N.~V.}\ \bibnamefont {Klenov}}, \bibinfo {author} {\bibfnamefont {S.~V.}\ \bibnamefont {Bakurskiy}}, \bibinfo {author} {\bibfnamefont {M.~Y.}\ \bibnamefont {Kupriyanov}}, \bibinfo {author} {\bibfnamefont {A.~L.}\ \bibnamefont {Gudkov}}, \ and\ \bibinfo {author} {\bibfnamefont {A.~S.}\ \bibnamefont {Sidorenko}},\ }\href {\doibase 10.3762/bjnano.8.269} {\bibfield  {journal} {\bibinfo  {journal} {Beilstein Journal of Nanotechnology}\ }\textbf {\bibinfo {volume} {8}},\ \bibinfo {pages} {2689} (\bibinfo {year} {2017})}\BibitemShut {NoStop}%
\bibitem [{\citenamefont {Likharev}\ and\ \citenamefont {Semenov}(1991)}]{Likharev1991}%
  \BibitemOpen
  \bibfield  {author} {\bibinfo {author} {\bibfnamefont {K.}~\bibnamefont {Likharev}}\ and\ \bibinfo {author} {\bibfnamefont {V.}~\bibnamefont {Semenov}},\ }\href {\doibase 10.1109/77.80745} {\bibfield  {journal} {\bibinfo  {journal} {IEEE Transactions on Applied Superconductivity}\ }\textbf {\bibinfo {volume} {1}},\ \bibinfo {pages} {3} (\bibinfo {year} {1991})}\BibitemShut {NoStop}%
\bibitem [{\citenamefont {Li}\ and\ \citenamefont {Yang}(2016)}]{Li2016}%
  \BibitemOpen
  \bibfield  {author} {\bibinfo {author} {\bibfnamefont {X.}~\bibnamefont {Li}}\ and\ \bibinfo {author} {\bibfnamefont {J.}~\bibnamefont {Yang}},\ }\href@noop {} {\bibfield  {journal} {\bibinfo  {journal} {National Science Review}\ }\textbf {\bibinfo {volume} {3}},\ \bibinfo {pages} {365} (\bibinfo {year} {2016})}\BibitemShut {NoStop}%
\bibitem [{\citenamefont {Shamiul~Alam}(2023)}]{Alam2023}%
  \BibitemOpen
  \bibfield  {author} {\bibinfo {author} {\bibfnamefont {S.~R. S. . A.~A.}\ \bibnamefont {Shamiul~Alam}, \bibfnamefont {Md~Shafayat~Hossain}},\ }\href {\doibase https://doi.org/10.1038/s41928-023-00930-2} {\bibfield  {journal} {\bibinfo  {journal} {Nature Electronics}\ }\textbf {\bibinfo {volume} {6}},\ \bibinfo {pages} {185–198} (\bibinfo {year} {2023})}\BibitemShut {NoStop}%
\bibitem [{\citenamefont {Natali}\ \emph {et~al.}(2010)\citenamefont {Natali}, \citenamefont {Plank}, \citenamefont {Galipaud}, \citenamefont {Ruck}, \citenamefont {Trodahl}, \citenamefont {Semond}, \citenamefont {Sorieul},\ and\ \citenamefont {Hirsch}}]{Natali2010}%
  \BibitemOpen
  \bibfield  {author} {\bibinfo {author} {\bibfnamefont {F.}~\bibnamefont {Natali}}, \bibinfo {author} {\bibfnamefont {N.~O.}\ \bibnamefont {Plank}}, \bibinfo {author} {\bibfnamefont {J.}~\bibnamefont {Galipaud}}, \bibinfo {author} {\bibfnamefont {B.~J.}\ \bibnamefont {Ruck}}, \bibinfo {author} {\bibfnamefont {H.~J.}\ \bibnamefont {Trodahl}}, \bibinfo {author} {\bibfnamefont {F.}~\bibnamefont {Semond}}, \bibinfo {author} {\bibfnamefont {S.}~\bibnamefont {Sorieul}}, \ and\ \bibinfo {author} {\bibfnamefont {L.}~\bibnamefont {Hirsch}},\ }\href {http://dx.doi.org/10.1016/j.jcrysgro.2010.09.030} {\bibfield  {journal} {\bibinfo  {journal} {Journal of Crystal Growth}\ }\textbf {\bibinfo {volume} {312}},\ \bibinfo {pages} {3583} (\bibinfo {year} {2010})}\BibitemShut {NoStop}%
\bibitem [{\citenamefont {Eick}\ \emph {et~al.}(1956)\citenamefont {Eick}, \citenamefont {Baenziger},\ and\ \citenamefont {Eyring}}]{Eick1956}%
  \BibitemOpen
  \bibfield  {author} {\bibinfo {author} {\bibfnamefont {H.~A.}\ \bibnamefont {Eick}}, \bibinfo {author} {\bibfnamefont {N.~C.}\ \bibnamefont {Baenziger}}, \ and\ \bibinfo {author} {\bibfnamefont {L.}~\bibnamefont {Eyring}},\ }\href@noop {} {\bibfield  {journal} {\bibinfo  {journal} {Journal of the American Chemical Society}\ }\textbf {\bibinfo {volume} {78}},\ \bibinfo {pages} {5987} (\bibinfo {year} {1956})}\BibitemShut {NoStop}%
\bibitem [{\citenamefont {Hulliger}(1978)}]{Hulliger1978}%
  \BibitemOpen
  \bibfield  {author} {\bibinfo {author} {\bibfnamefont {F.}~\bibnamefont {Hulliger}},\ }\href {\doibase https://doi.org/10.1016/0304-8853(78)90121-X} {\bibfield  {journal} {\bibinfo  {journal} {Journal of Magnetism and Magnetic Materials}\ }\textbf {\bibinfo {volume} {8}},\ \bibinfo {pages} {183 } (\bibinfo {year} {1978})}\BibitemShut {NoStop}%
\bibitem [{\citenamefont {Hulliger}(1979)}]{Hulliger1979}%
  \BibitemOpen
  \bibfield  {author} {\bibinfo {author} {\bibfnamefont {F.}~\bibnamefont {Hulliger}},\ }in\ \href {\doibase https://doi.org/10.1016/S0168-1273(79)04006-X} {\emph {\bibinfo {booktitle} {Non-Metallic Compounds - II}}},\ \bibinfo {series} {Handbook on the Physics and Chemistry of Rare Earths}, Vol.~\bibinfo {volume} {4}\ (\bibinfo  {publisher} {Elsevier},\ \bibinfo {year} {1979})\ pp.\ \bibinfo {pages} {153 -- 236}\BibitemShut {NoStop}%
\bibitem [{\citenamefont {Granville}\ \emph {et~al.}(2006)\citenamefont {Granville}, \citenamefont {Ruck}, \citenamefont {Budde}, \citenamefont {Koo}, \citenamefont {Pringle}, \citenamefont {Kuchler}, \citenamefont {Preston}, \citenamefont {Housden}, \citenamefont {Lund}, \citenamefont {Bittar}, \citenamefont {Williams},\ and\ \citenamefont {Trodahl}}]{Granville2006}%
  \BibitemOpen
  \bibfield  {author} {\bibinfo {author} {\bibfnamefont {S.}~\bibnamefont {Granville}}, \bibinfo {author} {\bibfnamefont {B.~J.}\ \bibnamefont {Ruck}}, \bibinfo {author} {\bibfnamefont {F.}~\bibnamefont {Budde}}, \bibinfo {author} {\bibfnamefont {A.}~\bibnamefont {Koo}}, \bibinfo {author} {\bibfnamefont {D.~J.}\ \bibnamefont {Pringle}}, \bibinfo {author} {\bibfnamefont {F.}~\bibnamefont {Kuchler}}, \bibinfo {author} {\bibfnamefont {A.~R.~H.}\ \bibnamefont {Preston}}, \bibinfo {author} {\bibfnamefont {D.~H.}\ \bibnamefont {Housden}}, \bibinfo {author} {\bibfnamefont {N.}~\bibnamefont {Lund}}, \bibinfo {author} {\bibfnamefont {A.}~\bibnamefont {Bittar}}, \bibinfo {author} {\bibfnamefont {G.~V.~M.}\ \bibnamefont {Williams}}, \ and\ \bibinfo {author} {\bibfnamefont {H.~J.}\ \bibnamefont {Trodahl}},\ }\href {\doibase 10.1103/PhysRevB.73.235335} {\bibfield  {journal} {\bibinfo  {journal} {Phys. Rev. B}\ }\textbf {\bibinfo {volume} {73}},\ \bibinfo {pages} {235335} (\bibinfo {year} {2006})}\BibitemShut {NoStop}%
\bibitem [{\citenamefont {Aerts}\ \emph {et~al.}(2004)\citenamefont {Aerts}, \citenamefont {Strange}, \citenamefont {Horne}, \citenamefont {Temmerman}, \citenamefont {Szotek},\ and\ \citenamefont {Svane}}]{Aerts2004}%
  \BibitemOpen
  \bibfield  {author} {\bibinfo {author} {\bibfnamefont {C.~M.}\ \bibnamefont {Aerts}}, \bibinfo {author} {\bibfnamefont {P.}~\bibnamefont {Strange}}, \bibinfo {author} {\bibfnamefont {M.}~\bibnamefont {Horne}}, \bibinfo {author} {\bibfnamefont {W.~M.}\ \bibnamefont {Temmerman}}, \bibinfo {author} {\bibfnamefont {Z.}~\bibnamefont {Szotek}}, \ and\ \bibinfo {author} {\bibfnamefont {A.}~\bibnamefont {Svane}},\ }\href {\doibase 10.1103/PhysRevB.69.045115} {\bibfield  {journal} {\bibinfo  {journal} {Phys. Rev. B}\ }\textbf {\bibinfo {volume} {69}},\ \bibinfo {pages} {045115} (\bibinfo {year} {2004})}\BibitemShut {NoStop}%
\bibitem [{\citenamefont {Anton}\ \emph {et~al.}(2016)\citenamefont {Anton}, \citenamefont {McNulty}, \citenamefont {Ruck}, \citenamefont {Suzuki}, \citenamefont {Mizumaki}, \citenamefont {Antonov}, \citenamefont {Quilty}, \citenamefont {Strickland},\ and\ \citenamefont {Trodahl}}]{Anton2016b}%
  \BibitemOpen
  \bibfield  {author} {\bibinfo {author} {\bibfnamefont {E.-M.}\ \bibnamefont {Anton}}, \bibinfo {author} {\bibfnamefont {J.~F.}\ \bibnamefont {McNulty}}, \bibinfo {author} {\bibfnamefont {B.~J.}\ \bibnamefont {Ruck}}, \bibinfo {author} {\bibfnamefont {M.}~\bibnamefont {Suzuki}}, \bibinfo {author} {\bibfnamefont {M.}~\bibnamefont {Mizumaki}}, \bibinfo {author} {\bibfnamefont {V.~N.}\ \bibnamefont {Antonov}}, \bibinfo {author} {\bibfnamefont {J.~W.}\ \bibnamefont {Quilty}}, \bibinfo {author} {\bibfnamefont {N.}~\bibnamefont {Strickland}}, \ and\ \bibinfo {author} {\bibfnamefont {H.~J.}\ \bibnamefont {Trodahl}},\ }\href {\doibase 10.1103/PhysRevB.93.064431} {\bibfield  {journal} {\bibinfo  {journal} {Phys. Rev. B}\ }\textbf {\bibinfo {volume} {93}},\ \bibinfo {pages} {064431} (\bibinfo {year} {2016})}\BibitemShut {NoStop}%
\bibitem [{\citenamefont {Holmes-Hewett}\ \emph {et~al.}(2020)\citenamefont {Holmes-Hewett}, \citenamefont {Pot}, \citenamefont {Buckley}, \citenamefont {Koo}, \citenamefont {Ruck}, \citenamefont {Natali}, \citenamefont {Shaib}, \citenamefont {Miller},\ and\ \citenamefont {Trodahl}}]{Holmes-Hewett2020}%
  \BibitemOpen
  \bibfield  {author} {\bibinfo {author} {\bibfnamefont {W.~F.}\ \bibnamefont {Holmes-Hewett}}, \bibinfo {author} {\bibfnamefont {C.}~\bibnamefont {Pot}}, \bibinfo {author} {\bibfnamefont {R.~G.}\ \bibnamefont {Buckley}}, \bibinfo {author} {\bibfnamefont {A.}~\bibnamefont {Koo}}, \bibinfo {author} {\bibfnamefont {B.~J.}\ \bibnamefont {Ruck}}, \bibinfo {author} {\bibfnamefont {F.}~\bibnamefont {Natali}}, \bibinfo {author} {\bibfnamefont {A.}~\bibnamefont {Shaib}}, \bibinfo {author} {\bibfnamefont {J.~D.}\ \bibnamefont {Miller}}, \ and\ \bibinfo {author} {\bibfnamefont {H.~J.}\ \bibnamefont {Trodahl}},\ }\href@noop {} {\bibfield  {journal} {\bibinfo  {journal} {Appl. Phys. Lett.}\ }\textbf {\bibinfo {volume} {117}},\ \bibinfo {pages} {222409} (\bibinfo {year} {2020})}\BibitemShut {NoStop}%
\bibitem [{\citenamefont {Holmes-Hewett}\ \emph {et~al.}(2023)\citenamefont {Holmes-Hewett}, \citenamefont {Koughnet}, \citenamefont {Miller}, \citenamefont {Trewick}, \citenamefont {Ruck}, \citenamefont {Trodahl},\ and\ \citenamefont {Buckley}}]{holmes-hewett2023}%
  \BibitemOpen
  \bibfield  {author} {\bibinfo {author} {\bibfnamefont {W.~F.}\ \bibnamefont {Holmes-Hewett}}, \bibinfo {author} {\bibfnamefont {K.~V.}\ \bibnamefont {Koughnet}}, \bibinfo {author} {\bibfnamefont {J.~D.}\ \bibnamefont {Miller}}, \bibinfo {author} {\bibfnamefont {E.~X.~M.}\ \bibnamefont {Trewick}}, \bibinfo {author} {\bibfnamefont {B.~J.}\ \bibnamefont {Ruck}}, \bibinfo {author} {\bibfnamefont {H.~J.}\ \bibnamefont {Trodahl}}, \ and\ \bibinfo {author} {\bibfnamefont {R.~G.}\ \bibnamefont {Buckley}},\ }\href@noop {} {\bibfield  {journal} {\bibinfo  {journal} {Sci Rep}\ }\textbf {\bibinfo {volume} {13}} (\bibinfo {year} {2023})}\BibitemShut {NoStop}%
\bibitem [{\citenamefont {Devese}\ \emph {et~al.}(2022)\citenamefont {Devese}, \citenamefont {Van~Koughnet}, \citenamefont {Buckley}, \citenamefont {Natali}, \citenamefont {Murmu}, \citenamefont {Anton}, \citenamefont {Ruck},\ and\ \citenamefont {Holmes-Hewett}}]{Devese2022}%
  \BibitemOpen
  \bibfield  {author} {\bibinfo {author} {\bibfnamefont {S.}~\bibnamefont {Devese}}, \bibinfo {author} {\bibfnamefont {K.}~\bibnamefont {Van~Koughnet}}, \bibinfo {author} {\bibfnamefont {R.~G.}\ \bibnamefont {Buckley}}, \bibinfo {author} {\bibfnamefont {F.}~\bibnamefont {Natali}}, \bibinfo {author} {\bibfnamefont {P.~P.}\ \bibnamefont {Murmu}}, \bibinfo {author} {\bibfnamefont {E.-M.}\ \bibnamefont {Anton}}, \bibinfo {author} {\bibfnamefont {B.~J.}\ \bibnamefont {Ruck}}, \ and\ \bibinfo {author} {\bibfnamefont {W.~F.}\ \bibnamefont {Holmes-Hewett}},\ }\href@noop {} {\bibfield  {journal} {\bibinfo  {journal} {AIP Advances}\ }\textbf {\bibinfo {volume} {12}},\ \bibinfo {pages} {035108} (\bibinfo {year} {2022})}\BibitemShut {NoStop}%
\bibitem [{\citenamefont {Ludbrook}\ \emph {et~al.}(2009)\citenamefont {Ludbrook}, \citenamefont {Farrell}, \citenamefont {Kübel}, \citenamefont {Ruck}, \citenamefont {Preston}, \citenamefont {Trodahl}, \citenamefont {Ranno}, \citenamefont {Reeves},\ and\ \citenamefont {Durbin}}]{Ludbrook2009}%
  \BibitemOpen
  \bibfield  {author} {\bibinfo {author} {\bibfnamefont {B.}~\bibnamefont {Ludbrook}}, \bibinfo {author} {\bibfnamefont {I.}~\bibnamefont {Farrell}}, \bibinfo {author} {\bibfnamefont {M.}~\bibnamefont {Kübel}}, \bibinfo {author} {\bibfnamefont {B.}~\bibnamefont {Ruck}}, \bibinfo {author} {\bibfnamefont {A.}~\bibnamefont {Preston}}, \bibinfo {author} {\bibfnamefont {J.}~\bibnamefont {Trodahl}}, \bibinfo {author} {\bibfnamefont {L.}~\bibnamefont {Ranno}}, \bibinfo {author} {\bibfnamefont {R.}~\bibnamefont {Reeves}}, \ and\ \bibinfo {author} {\bibfnamefont {S.}~\bibnamefont {Durbin}},\ }\href {\doibase 10.1063/1.3211290} {\bibfield  {journal} {\bibinfo  {journal} {Journal of Applied Physics}\ }\textbf {\bibinfo {volume} {106}},\ \bibinfo {pages} {063910} (\bibinfo {year} {2009})}\BibitemShut {NoStop}%
\bibitem [{\citenamefont {Trodahl}\ \emph {et~al.}(2007)\citenamefont {Trodahl}, \citenamefont {Preston}, \citenamefont {Zhong}, \citenamefont {Ruck}, \citenamefont {Strickland}, \citenamefont {Mitra},\ and\ \citenamefont {Lambrecht}}]{Trodahl2007}%
  \BibitemOpen
  \bibfield  {author} {\bibinfo {author} {\bibfnamefont {H.~J.}\ \bibnamefont {Trodahl}}, \bibinfo {author} {\bibfnamefont {A.~R.~H.}\ \bibnamefont {Preston}}, \bibinfo {author} {\bibfnamefont {J.}~\bibnamefont {Zhong}}, \bibinfo {author} {\bibfnamefont {B.~J.}\ \bibnamefont {Ruck}}, \bibinfo {author} {\bibfnamefont {N.~M.}\ \bibnamefont {Strickland}}, \bibinfo {author} {\bibfnamefont {C.}~\bibnamefont {Mitra}}, \ and\ \bibinfo {author} {\bibfnamefont {W.~R.~L.}\ \bibnamefont {Lambrecht}},\ }\href@noop {} {\bibfield  {journal} {\bibinfo  {journal} {Phys. Rev. B}\ }\textbf {\bibinfo {volume} {76}},\ \bibinfo {pages} {085211} (\bibinfo {year} {2007})}\BibitemShut {NoStop}%
\bibitem [{\citenamefont {Larson}\ and\ \citenamefont {Lambrecht}(2006)}]{Larson2006}%
  \BibitemOpen
  \bibfield  {author} {\bibinfo {author} {\bibfnamefont {P.}~\bibnamefont {Larson}}\ and\ \bibinfo {author} {\bibfnamefont {W.~R.~L.}\ \bibnamefont {Lambrecht}},\ }\href {\doibase 10.1103/PhysRevB.74.085108} {\bibfield  {journal} {\bibinfo  {journal} {Phys. Rev. B}\ }\textbf {\bibinfo {volume} {74}},\ \bibinfo {pages} {085108} (\bibinfo {year} {2006})}\BibitemShut {NoStop}%
\bibitem [{\citenamefont {Larson}\ \emph {et~al.}(2007)\citenamefont {Larson}, \citenamefont {Lambrecht}, \citenamefont {Chantis},\ and\ \citenamefont {vanSchilfgaarde}}]{Larson2007}%
  \BibitemOpen
  \bibfield  {author} {\bibinfo {author} {\bibfnamefont {P.}~\bibnamefont {Larson}}, \bibinfo {author} {\bibfnamefont {W.~R.~L.}\ \bibnamefont {Lambrecht}}, \bibinfo {author} {\bibfnamefont {A.}~\bibnamefont {Chantis}}, \ and\ \bibinfo {author} {\bibfnamefont {M.}~\bibnamefont {vanSchilfgaarde}},\ }\href@noop {} {\bibfield  {journal} {\bibinfo  {journal} {Phys. Rev. B}\ }\textbf {\bibinfo {volume} {75}},\ \bibinfo {pages} {045114} (\bibinfo {year} {2007})}\BibitemShut {NoStop}%
\bibitem [{\citenamefont {Preston}\ \emph {et~al.}(2010)\citenamefont {Preston}, \citenamefont {Ruck}, \citenamefont {Lambrecht}, \citenamefont {Piper}, \citenamefont {Downes}, \citenamefont {Smith},\ and\ \citenamefont {Trodahl}}]{Preston2010a}%
  \BibitemOpen
  \bibfield  {author} {\bibinfo {author} {\bibfnamefont {A.~R.~H.}\ \bibnamefont {Preston}}, \bibinfo {author} {\bibfnamefont {B.~J.}\ \bibnamefont {Ruck}}, \bibinfo {author} {\bibfnamefont {W.~R.~L.}\ \bibnamefont {Lambrecht}}, \bibinfo {author} {\bibfnamefont {L.~F.~J.}\ \bibnamefont {Piper}}, \bibinfo {author} {\bibfnamefont {J.~E.}\ \bibnamefont {Downes}}, \bibinfo {author} {\bibfnamefont {K.~E.}\ \bibnamefont {Smith}}, \ and\ \bibinfo {author} {\bibfnamefont {H.~J.}\ \bibnamefont {Trodahl}},\ }\href {\doibase 10.1063/1.3291057} {\bibfield  {journal} {\bibinfo  {journal} {Appl. Phys. Lett.}\ }\textbf {\bibinfo {volume} {96}},\ \bibinfo {pages} {032101} (\bibinfo {year} {2010})}\BibitemShut {NoStop}%
\bibitem [{\citenamefont {Vilela}\ \emph {et~al.}(2024)\citenamefont {Vilela}, \citenamefont {Stephen}, \citenamefont {Gratens}, \citenamefont {Galgano}, \citenamefont {Hou}, \citenamefont {Takamura}, \citenamefont {Heiman}, \citenamefont {Henriques}, \citenamefont {Berera},\ and\ \citenamefont {Moodera}}]{Vilela2024}%
  \BibitemOpen
  \bibfield  {author} {\bibinfo {author} {\bibfnamefont {G.~L.~S.}\ \bibnamefont {Vilela}}, \bibinfo {author} {\bibfnamefont {G.~M.}\ \bibnamefont {Stephen}}, \bibinfo {author} {\bibfnamefont {X.}~\bibnamefont {Gratens}}, \bibinfo {author} {\bibfnamefont {G.~D.}\ \bibnamefont {Galgano}}, \bibinfo {author} {\bibfnamefont {Y.}~\bibnamefont {Hou}}, \bibinfo {author} {\bibfnamefont {Y.}~\bibnamefont {Takamura}}, \bibinfo {author} {\bibfnamefont {D.}~\bibnamefont {Heiman}}, \bibinfo {author} {\bibfnamefont {A.~B.}\ \bibnamefont {Henriques}}, \bibinfo {author} {\bibfnamefont {G.}~\bibnamefont {Berera}}, \ and\ \bibinfo {author} {\bibfnamefont {J.~S.}\ \bibnamefont {Moodera}},\ }\href {\doibase 10.1103/PhysRevB.109.L060401} {\bibfield  {journal} {\bibinfo  {journal} {Phys. Rev. B}\ }\textbf {\bibinfo {volume} {109}},\ \bibinfo {pages} {L060401} (\bibinfo {year} {2024})}\BibitemShut {NoStop}%
\bibitem [{\citenamefont {Azeem}(2016)}]{Azeem2016}%
  \BibitemOpen
  \bibfield  {author} {\bibinfo {author} {\bibfnamefont {M.}~\bibnamefont {Azeem}},\ }\href {\doibase 10.1088/0256-307x/33/2/027501} {\bibfield  {journal} {\bibinfo  {journal} {Chin. Phys. Lett.}\ }\textbf {\bibinfo {volume} {33}},\ \bibinfo {pages} {027501} (\bibinfo {year} {2016})}\BibitemShut {NoStop}%
\bibitem [{\citenamefont {Yoshitomi}\ \emph {et~al.}(2011)\citenamefont {Yoshitomi}, \citenamefont {Kitayama}, \citenamefont {Kita}, \citenamefont {Wada}, \citenamefont {Fujisawa}, \citenamefont {Ohta},\ and\ \citenamefont {Sakurai}}]{Yoshitomi2011}%
  \BibitemOpen
  \bibfield  {author} {\bibinfo {author} {\bibfnamefont {H.}~\bibnamefont {Yoshitomi}}, \bibinfo {author} {\bibfnamefont {S.}~\bibnamefont {Kitayama}}, \bibinfo {author} {\bibfnamefont {T.}~\bibnamefont {Kita}}, \bibinfo {author} {\bibfnamefont {O.}~\bibnamefont {Wada}}, \bibinfo {author} {\bibfnamefont {M.}~\bibnamefont {Fujisawa}}, \bibinfo {author} {\bibfnamefont {H.}~\bibnamefont {Ohta}}, \ and\ \bibinfo {author} {\bibfnamefont {T.}~\bibnamefont {Sakurai}},\ }\href@noop {} {\bibfield  {journal} {\bibinfo  {journal} {Phys. Rev. B}\ }\textbf {\bibinfo {volume} {83}},\ \bibinfo {pages} {155202} (\bibinfo {year} {2011})}\BibitemShut {NoStop}%
\bibitem [{\citenamefont {Vidyasagar}\ \emph {et~al.}(2012)\citenamefont {Vidyasagar}, \citenamefont {Kitayama}, \citenamefont {Yoshitomi}, \citenamefont {Kita}, \citenamefont {Sakurai},\ and\ \citenamefont {Ohta}}]{Vidyasagar2012}%
  \BibitemOpen
  \bibfield  {author} {\bibinfo {author} {\bibfnamefont {R.}~\bibnamefont {Vidyasagar}}, \bibinfo {author} {\bibfnamefont {S.}~\bibnamefont {Kitayama}}, \bibinfo {author} {\bibfnamefont {H.}~\bibnamefont {Yoshitomi}}, \bibinfo {author} {\bibfnamefont {T.}~\bibnamefont {Kita}}, \bibinfo {author} {\bibfnamefont {T.}~\bibnamefont {Sakurai}}, \ and\ \bibinfo {author} {\bibfnamefont {H.}~\bibnamefont {Ohta}},\ }\href {\doibase 10.1063/1.4727903} {\bibfield  {journal} {\bibinfo  {journal} {Appl. Phys. Lett.}\ }\textbf {\bibinfo {volume} {100}},\ \bibinfo {pages} {232410} (\bibinfo {year} {2012})}\BibitemShut {NoStop}%
\bibitem [{\citenamefont {Maity}\ \emph {et~al.}(2018)\citenamefont {Maity}, \citenamefont {Trodahl}, \citenamefont {Natali}, \citenamefont {Ruck},\ and\ \citenamefont {V\'ezian}}]{Maity2018}%
  \BibitemOpen
  \bibfield  {author} {\bibinfo {author} {\bibfnamefont {T.}~\bibnamefont {Maity}}, \bibinfo {author} {\bibfnamefont {H.~J.}\ \bibnamefont {Trodahl}}, \bibinfo {author} {\bibfnamefont {F.}~\bibnamefont {Natali}}, \bibinfo {author} {\bibfnamefont {B.~J.}\ \bibnamefont {Ruck}}, \ and\ \bibinfo {author} {\bibfnamefont {S.}~\bibnamefont {V\'ezian}},\ }\href {\doibase 10.1103/PhysRevMaterials.2.014405} {\bibfield  {journal} {\bibinfo  {journal} {Phys. Rev. Materials}\ }\textbf {\bibinfo {volume} {2}},\ \bibinfo {pages} {014405} (\bibinfo {year} {2018})}\BibitemShut {NoStop}%
\bibitem [{\citenamefont {Punya}\ \emph {et~al.}(2011)\citenamefont {Punya}, \citenamefont {Cheiwchanchamnangij}, \citenamefont {Thiess},\ and\ \citenamefont {Lambrecht}}]{Punya2011}%
  \BibitemOpen
  \bibfield  {author} {\bibinfo {author} {\bibfnamefont {A.}~\bibnamefont {Punya}}, \bibinfo {author} {\bibfnamefont {T.}~\bibnamefont {Cheiwchanchamnangij}}, \bibinfo {author} {\bibfnamefont {A.}~\bibnamefont {Thiess}}, \ and\ \bibinfo {author} {\bibfnamefont {W.}~\bibnamefont {Lambrecht}},\ }\href@noop {} {\bibfield  {journal} {\bibinfo  {journal} {MRS Proceedings}\ }\textbf {\bibinfo {volume} {1290}} (\bibinfo {year} {2011})}\BibitemShut {NoStop}%
\bibitem [{\citenamefont {Trodahl}\ \emph {et~al.}(2017)\citenamefont {Trodahl}, \citenamefont {Natali}, \citenamefont {Ruck},\ and\ \citenamefont {Lambrecht}}]{Trodahl2017}%
  \BibitemOpen
  \bibfield  {author} {\bibinfo {author} {\bibfnamefont {H.~J.}\ \bibnamefont {Trodahl}}, \bibinfo {author} {\bibfnamefont {F.}~\bibnamefont {Natali}}, \bibinfo {author} {\bibfnamefont {B.~J.}\ \bibnamefont {Ruck}}, \ and\ \bibinfo {author} {\bibfnamefont {W.~R.~L.}\ \bibnamefont {Lambrecht}},\ }\href@noop {} {\bibfield  {journal} {\bibinfo  {journal} {Phys. Rev. B}\ }\textbf {\bibinfo {volume} {96}},\ \bibinfo {pages} {115309} (\bibinfo {year} {2017})}\BibitemShut {NoStop}%
\bibitem [{\citenamefont {Birge}\ and\ \citenamefont {Satchell}(2024)}]{Birge2024}%
  \BibitemOpen
  \bibfield  {author} {\bibinfo {author} {\bibfnamefont {N.~O.}\ \bibnamefont {Birge}}\ and\ \bibinfo {author} {\bibfnamefont {N.}~\bibnamefont {Satchell}},\ }\href@noop {} {\bibfield  {journal} {\bibinfo  {journal} {APL Materials}\ }\textbf {\bibinfo {volume} {12}},\ \bibinfo {pages} {041105} (\bibinfo {year} {2024})}\BibitemShut {NoStop}%
\bibitem [{\citenamefont {Vernik}\ \emph {et~al.}(2013)\citenamefont {Vernik}, \citenamefont {Bol'ginov}, \citenamefont {Bakurskiy}, \citenamefont {Golubov}, \citenamefont {Kupriyanov}, \citenamefont {Ryazanov},\ and\ \citenamefont {Mukhanov}}]{Vernik2013}%
  \BibitemOpen
  \bibfield  {author} {\bibinfo {author} {\bibfnamefont {I.~V.}\ \bibnamefont {Vernik}}, \bibinfo {author} {\bibfnamefont {V.~V.}\ \bibnamefont {Bol'ginov}}, \bibinfo {author} {\bibfnamefont {S.~V.}\ \bibnamefont {Bakurskiy}}, \bibinfo {author} {\bibfnamefont {A.~A.}\ \bibnamefont {Golubov}}, \bibinfo {author} {\bibfnamefont {M.~Y.}\ \bibnamefont {Kupriyanov}}, \bibinfo {author} {\bibfnamefont {V.~V.}\ \bibnamefont {Ryazanov}}, \ and\ \bibinfo {author} {\bibfnamefont {O.~A.}\ \bibnamefont {Mukhanov}},\ }\href@noop {} {\bibfield  {journal} {\bibinfo  {journal} {IEEE Transactions on Applied Superconductivity}\ }\textbf {\bibinfo {volume} {23}},\ \bibinfo {pages} {1701208} (\bibinfo {year} {2013})}\BibitemShut {NoStop}%
\bibitem [{\citenamefont {Senapati}\ \emph {et~al.}(2011)\citenamefont {Senapati}, \citenamefont {Blamire},\ and\ \citenamefont {Barber}}]{Senapati2011}%
  \BibitemOpen
  \bibfield  {author} {\bibinfo {author} {\bibfnamefont {K.}~\bibnamefont {Senapati}}, \bibinfo {author} {\bibfnamefont {M.~G.}\ \bibnamefont {Blamire}}, \ and\ \bibinfo {author} {\bibfnamefont {Z.~H.}\ \bibnamefont {Barber}},\ }\href@noop {} {\bibfield  {journal} {\bibinfo  {journal} {Nat. Mater.}\ }\textbf {\bibinfo {volume} {10}},\ \bibinfo {pages} {849} (\bibinfo {year} {2011})}\BibitemShut {NoStop}%
\bibitem [{\citenamefont {Massarotti}\ \emph {et~al.}(2015)\citenamefont {Massarotti}, \citenamefont {A.Pal}, \citenamefont {Rotoli}, \citenamefont {Longobardi}, \citenamefont {Blamire},\ and\ \citenamefont {Tafuri}}]{Massarotti2015}%
  \BibitemOpen
  \bibfield  {author} {\bibinfo {author} {\bibfnamefont {D.}~\bibnamefont {Massarotti}}, \bibinfo {author} {\bibnamefont {A.Pal}}, \bibinfo {author} {\bibfnamefont {G.}~\bibnamefont {Rotoli}}, \bibinfo {author} {\bibfnamefont {L.}~\bibnamefont {Longobardi}}, \bibinfo {author} {\bibfnamefont {M.}~\bibnamefont {Blamire}}, \ and\ \bibinfo {author} {\bibfnamefont {F.}~\bibnamefont {Tafuri}},\ }\href@noop {} {\bibfield  {journal} {\bibinfo  {journal} {Nature Communications}\ }\textbf {\bibinfo {volume} {6}},\ \bibinfo {pages} {7376} (\bibinfo {year} {2015})}\BibitemShut {NoStop}%
\bibitem [{\citenamefont {Ahmad}\ \emph {et~al.}(2020)\citenamefont {Ahmad}, \citenamefont {Caruso}, \citenamefont {Pal}, \citenamefont {Rotoli}, \citenamefont {Pepe}, \citenamefont {Blamire}, \citenamefont {Tafuri},\ and\ \citenamefont {Massarotti}}]{Ahmad2020}%
  \BibitemOpen
  \bibfield  {author} {\bibinfo {author} {\bibfnamefont {H.}~\bibnamefont {Ahmad}}, \bibinfo {author} {\bibfnamefont {R.}~\bibnamefont {Caruso}}, \bibinfo {author} {\bibfnamefont {A.}~\bibnamefont {Pal}}, \bibinfo {author} {\bibfnamefont {G.}~\bibnamefont {Rotoli}}, \bibinfo {author} {\bibfnamefont {G.}~\bibnamefont {Pepe}}, \bibinfo {author} {\bibfnamefont {M.}~\bibnamefont {Blamire}}, \bibinfo {author} {\bibfnamefont {F.}~\bibnamefont {Tafuri}}, \ and\ \bibinfo {author} {\bibfnamefont {D.}~\bibnamefont {Massarotti}},\ }\href {\doibase 10.1103/PhysRevApplied.13.014017} {\bibfield  {journal} {\bibinfo  {journal} {Phys. Rev. Appl.}\ }\textbf {\bibinfo {volume} {13}},\ \bibinfo {pages} {014017} (\bibinfo {year} {2020})}\BibitemShut {NoStop}%
\bibitem [{\citenamefont {Sharma}\ \emph {et~al.}(2023)\citenamefont {Sharma}, \citenamefont {Banerjee}, \citenamefont {Dutta}, \citenamefont {Singhal}, \citenamefont {Banerjee}, \citenamefont {Pal},\ and\ \citenamefont {Pal}}]{Sharma2023}%
  \BibitemOpen
  \bibfield  {author} {\bibinfo {author} {\bibfnamefont {P.~K.}\ \bibnamefont {Sharma}}, \bibinfo {author} {\bibfnamefont {S.}~\bibnamefont {Banerjee}}, \bibinfo {author} {\bibfnamefont {B.}~\bibnamefont {Dutta}}, \bibinfo {author} {\bibfnamefont {V.}~\bibnamefont {Singhal}}, \bibinfo {author} {\bibfnamefont {P.}~\bibnamefont {Banerjee}}, \bibinfo {author} {\bibfnamefont {H.~K.}\ \bibnamefont {Pal}}, \ and\ \bibinfo {author} {\bibfnamefont {A.}~\bibnamefont {Pal}},\ }\href@noop {} {\  (\bibinfo {year} {2023})},\ \Eprint {http://arxiv.org/abs/2312.04650} {arXiv:2312.04650 [cond-mat.supr-con]} \BibitemShut {NoStop}%
\bibitem [{\citenamefont {Cascales}\ \emph {et~al.}(2019)\citenamefont {Cascales}, \citenamefont {Takamura}, \citenamefont {Stephen}, \citenamefont {Heiman}, \citenamefont {Bergeret},\ and\ \citenamefont {Moodera}}]{Cascales2019}%
  \BibitemOpen
  \bibfield  {author} {\bibinfo {author} {\bibfnamefont {J.~P.}\ \bibnamefont {Cascales}}, \bibinfo {author} {\bibfnamefont {Y.}~\bibnamefont {Takamura}}, \bibinfo {author} {\bibfnamefont {G.~M.}\ \bibnamefont {Stephen}}, \bibinfo {author} {\bibfnamefont {D.}~\bibnamefont {Heiman}}, \bibinfo {author} {\bibfnamefont {F.~S.}\ \bibnamefont {Bergeret}}, \ and\ \bibinfo {author} {\bibfnamefont {J.~S.}\ \bibnamefont {Moodera}},\ }\href@noop {} {\bibfield  {journal} {\bibinfo  {journal} {Applied Physics Letters}\ }\textbf {\bibinfo {volume} {114}},\ \bibinfo {pages} {022601} (\bibinfo {year} {2019})}\BibitemShut {NoStop}%
\bibitem [{\citenamefont {Pot}\ \emph {et~al.}(2023)\citenamefont {Pot}, \citenamefont {Holmes-Hewett}, \citenamefont {Anton}, \citenamefont {Miller}, \citenamefont {Ruck},\ and\ \citenamefont {Trodahl}}]{pot2023}%
  \BibitemOpen
  \bibfield  {author} {\bibinfo {author} {\bibfnamefont {C.}~\bibnamefont {Pot}}, \bibinfo {author} {\bibfnamefont {W.~F.}\ \bibnamefont {Holmes-Hewett}}, \bibinfo {author} {\bibfnamefont {E.-M.}\ \bibnamefont {Anton}}, \bibinfo {author} {\bibfnamefont {J.~D.}\ \bibnamefont {Miller}}, \bibinfo {author} {\bibfnamefont {B.~J.}\ \bibnamefont {Ruck}}, \ and\ \bibinfo {author} {\bibfnamefont {H.~J.}\ \bibnamefont {Trodahl}},\ }\href@noop {} {\bibfield  {journal} {\bibinfo  {journal} {Appl. Phys. Lett.}\ }\textbf {\bibinfo {volume} {13}} (\bibinfo {year} {2023})}\BibitemShut {NoStop}%
\bibitem [{\citenamefont {Ahmad}\ \emph {et~al.}(2022)\citenamefont {Ahmad}, \citenamefont {Brosco}, \citenamefont {Miano}, \citenamefont {Di~Palma}, \citenamefont {Arzeo}, \citenamefont {Montemurro}, \citenamefont {Lucignano}, \citenamefont {Pepe}, \citenamefont {Tafuri}, \citenamefont {Fazio},\ and\ \citenamefont {Massarotti}}]{Ahmad2022b}%
  \BibitemOpen
  \bibfield  {author} {\bibinfo {author} {\bibfnamefont {H.~G.}\ \bibnamefont {Ahmad}}, \bibinfo {author} {\bibfnamefont {V.}~\bibnamefont {Brosco}}, \bibinfo {author} {\bibfnamefont {A.}~\bibnamefont {Miano}}, \bibinfo {author} {\bibfnamefont {L.}~\bibnamefont {Di~Palma}}, \bibinfo {author} {\bibfnamefont {M.}~\bibnamefont {Arzeo}}, \bibinfo {author} {\bibfnamefont {D.}~\bibnamefont {Montemurro}}, \bibinfo {author} {\bibfnamefont {P.}~\bibnamefont {Lucignano}}, \bibinfo {author} {\bibfnamefont {G.~P.}\ \bibnamefont {Pepe}}, \bibinfo {author} {\bibfnamefont {F.}~\bibnamefont {Tafuri}}, \bibinfo {author} {\bibfnamefont {R.}~\bibnamefont {Fazio}}, \ and\ \bibinfo {author} {\bibfnamefont {D.}~\bibnamefont {Massarotti}},\ }\href {\doibase 10.1103/PhysRevB.105.214522} {\bibfield  {journal} {\bibinfo  {journal} {Phys. Rev. B}\ }\textbf {\bibinfo {volume} {105}},\ \bibinfo {pages} {214522} (\bibinfo {year} {2022})}\BibitemShut {NoStop}%
\bibitem [{\citenamefont {Mitra}\ and\ \citenamefont {Lambrecht}(2008)}]{Mitra2008}%
  \BibitemOpen
  \bibfield  {author} {\bibinfo {author} {\bibfnamefont {C.}~\bibnamefont {Mitra}}\ and\ \bibinfo {author} {\bibfnamefont {W.~R.~L.}\ \bibnamefont {Lambrecht}},\ }\href {\doibase 10.1103/PhysRevB.78.195203} {\bibfield  {journal} {\bibinfo  {journal} {Phys. Rev. B}\ }\textbf {\bibinfo {volume} {78}},\ \bibinfo {pages} {195203} (\bibinfo {year} {2008})}\BibitemShut {NoStop}%
\bibitem [{\citenamefont {Giannozzi}\ \emph {et~al.}(2009)\citenamefont {Giannozzi} \emph {et~al.}}]{QE}%
  \BibitemOpen
  \bibfield  {author} {\bibinfo {author} {\bibfnamefont {P.}~\bibnamefont {Giannozzi}} \emph {et~al.},\ }\href {\doibase 10.1088/0953-8984/21/39/395502} {\bibfield  {journal} {\bibinfo  {journal} {J. Phys.: Condens. Matter}\ }\textbf {\bibinfo {volume} {21}},\ \bibinfo {pages} {395502} (\bibinfo {year} {2009})}\BibitemShut {NoStop}%
\bibitem [{\citenamefont {Cococcioni}\ and\ \citenamefont {de~Gironcoli}(2005)}]{Cococcioni2005}%
  \BibitemOpen
  \bibfield  {author} {\bibinfo {author} {\bibfnamefont {M.}~\bibnamefont {Cococcioni}}\ and\ \bibinfo {author} {\bibfnamefont {S.}~\bibnamefont {de~Gironcoli}},\ }\href {\doibase 10.1103/PhysRevB.71.035105} {\bibfield  {journal} {\bibinfo  {journal} {Phys. Rev. B}\ }\textbf {\bibinfo {volume} {71}},\ \bibinfo {pages} {035105} (\bibinfo {year} {2005})}\BibitemShut {NoStop}%
\bibitem [{\citenamefont {Topsakal}\ and\ \citenamefont {Wentzcovitch}(2014)}]{Topsakal2014}%
  \BibitemOpen
  \bibfield  {author} {\bibinfo {author} {\bibfnamefont {M.}~\bibnamefont {Topsakal}}\ and\ \bibinfo {author} {\bibfnamefont {R.}~\bibnamefont {Wentzcovitch}},\ }\href {\doibase https://doi.org/10.1016/j.commatsci.2014.07.030} {\bibfield  {journal} {\bibinfo  {journal} {Comput. Mater. Sci.}\ }\textbf {\bibinfo {volume} {95}},\ \bibinfo {pages} {263 } (\bibinfo {year} {2014})}\BibitemShut {NoStop}%
\bibitem [{\citenamefont {Mostofi}\ \emph {et~al.}(2014)\citenamefont {Mostofi}, \citenamefont {Yates}, \citenamefont {Pizzi}, \citenamefont {Lee}, \citenamefont {Souza}, \citenamefont {Vanderbilt},\ and\ \citenamefont {Marzari}}]{Mostofi2014}%
  \BibitemOpen
  \bibfield  {author} {\bibinfo {author} {\bibfnamefont {A.~A.}\ \bibnamefont {Mostofi}}, \bibinfo {author} {\bibfnamefont {J.~R.}\ \bibnamefont {Yates}}, \bibinfo {author} {\bibfnamefont {G.}~\bibnamefont {Pizzi}}, \bibinfo {author} {\bibfnamefont {Y.-S.}\ \bibnamefont {Lee}}, \bibinfo {author} {\bibfnamefont {I.}~\bibnamefont {Souza}}, \bibinfo {author} {\bibfnamefont {D.}~\bibnamefont {Vanderbilt}}, \ and\ \bibinfo {author} {\bibfnamefont {N.}~\bibnamefont {Marzari}},\ }\href {\doibase https://doi.org/10.1016/j.cpc.2014.05.003} {\bibfield  {journal} {\bibinfo  {journal} {Computer Physics Communications}\ }\textbf {\bibinfo {volume} {185}},\ \bibinfo {pages} {2309} (\bibinfo {year} {2014})}\BibitemShut {NoStop}%
\bibitem [{\citenamefont {Marzari}\ and\ \citenamefont {Vanderbilt}(1997)}]{Marzari1997}%
  \BibitemOpen
  \bibfield  {author} {\bibinfo {author} {\bibfnamefont {N.}~\bibnamefont {Marzari}}\ and\ \bibinfo {author} {\bibfnamefont {D.}~\bibnamefont {Vanderbilt}},\ }\href {\doibase 10.1103/PhysRevB.56.12847} {\bibfield  {journal} {\bibinfo  {journal} {Phys. Rev. B}\ }\textbf {\bibinfo {volume} {56}},\ \bibinfo {pages} {12847} (\bibinfo {year} {1997})}\BibitemShut {NoStop}%
\bibitem [{\citenamefont {Souza}\ \emph {et~al.}(2001)\citenamefont {Souza}, \citenamefont {Marzari},\ and\ \citenamefont {Vanderbilt}}]{Souza2001}%
  \BibitemOpen
  \bibfield  {author} {\bibinfo {author} {\bibfnamefont {I.}~\bibnamefont {Souza}}, \bibinfo {author} {\bibfnamefont {N.}~\bibnamefont {Marzari}}, \ and\ \bibinfo {author} {\bibfnamefont {D.}~\bibnamefont {Vanderbilt}},\ }\href {\doibase 10.1103/PhysRevB.65.035109} {\bibfield  {journal} {\bibinfo  {journal} {Phys. Rev. B}\ }\textbf {\bibinfo {volume} {65}},\ \bibinfo {pages} {035109} (\bibinfo {year} {2001})}\BibitemShut {NoStop}%
\bibitem [{\citenamefont {Holmes-Hewett}(2021)}]{Holmes-Hewett2021}%
  \BibitemOpen
  \bibfield  {author} {\bibinfo {author} {\bibfnamefont {W.~F.}\ \bibnamefont {Holmes-Hewett}},\ }\href {https://link.aps.org/doi/10.1103/PhysRevB.104.075124} {\bibfield  {journal} {\bibinfo  {journal} {Phys. Rev. B}\ }\textbf {\bibinfo {volume} {104}},\ \bibinfo {pages} {075124} (\bibinfo {year} {2021})}\BibitemShut {NoStop}%
\bibitem [{\citenamefont {Ehrenreich}\ \emph {et~al.}(1963)\citenamefont {Ehrenreich}, \citenamefont {Philipp},\ and\ \citenamefont {Segall}}]{Ehrenreich1963}%
  \BibitemOpen
  \bibfield  {author} {\bibinfo {author} {\bibfnamefont {H.}~\bibnamefont {Ehrenreich}}, \bibinfo {author} {\bibfnamefont {H.~R.}\ \bibnamefont {Philipp}}, \ and\ \bibinfo {author} {\bibfnamefont {B.}~\bibnamefont {Segall}},\ }\href {\doibase 10.1103/PhysRev.132.1918} {\bibfield  {journal} {\bibinfo  {journal} {Phys. Rev.}\ }\textbf {\bibinfo {volume} {132}},\ \bibinfo {pages} {1918} (\bibinfo {year} {1963})}\BibitemShut {NoStop}%
\bibitem [{\citenamefont {Kuzmenko}(2005)}]{RefFit}%
  \BibitemOpen
  \bibfield  {author} {\bibinfo {author} {\bibfnamefont {A.~B.}\ \bibnamefont {Kuzmenko}},\ }\href@noop {} {\bibfield  {journal} {\bibinfo  {journal} {Rev. Sci. Instrum.}\ }\textbf {\bibinfo {volume} {76}},\ \bibinfo {pages} {083108} (\bibinfo {year} {2005})}\BibitemShut {NoStop}%
\bibitem [{\citenamefont {Lee}\ \emph {et~al.}(2015)\citenamefont {Lee}, \citenamefont {Warring}, \citenamefont {V{\'{e}}zian}, \citenamefont {Damilano}, \citenamefont {Granville}, \citenamefont {{Al Khalfioui}}, \citenamefont {Cordier}, \citenamefont {Trodahl}, \citenamefont {Ruck},\ and\ \citenamefont {Natali}}]{Lee2015}%
  \BibitemOpen
  \bibfield  {author} {\bibinfo {author} {\bibfnamefont {C.~M.}\ \bibnamefont {Lee}}, \bibinfo {author} {\bibfnamefont {H.}~\bibnamefont {Warring}}, \bibinfo {author} {\bibfnamefont {S.}~\bibnamefont {V{\'{e}}zian}}, \bibinfo {author} {\bibfnamefont {B.}~\bibnamefont {Damilano}}, \bibinfo {author} {\bibfnamefont {S.}~\bibnamefont {Granville}}, \bibinfo {author} {\bibfnamefont {M.}~\bibnamefont {{Al Khalfioui}}}, \bibinfo {author} {\bibfnamefont {Y.}~\bibnamefont {Cordier}}, \bibinfo {author} {\bibfnamefont {H.~J.}\ \bibnamefont {Trodahl}}, \bibinfo {author} {\bibfnamefont {B.~J.}\ \bibnamefont {Ruck}}, \ and\ \bibinfo {author} {\bibfnamefont {F.}~\bibnamefont {Natali}},\ }\href@noop {} {\bibfield  {journal} {\bibinfo  {journal} {Appl. Phys. Lett.}\ }\textbf {\bibinfo {volume} {106}},\ \bibinfo {pages} {022401} (\bibinfo {year} {2015})}\BibitemShut {NoStop}%
\bibitem [{\citenamefont {Holmes-Hewett}\ \emph {et~al.}(2019)\citenamefont {Holmes-Hewett}, \citenamefont {Buckley}, \citenamefont {Ruck}, \citenamefont {Natali},\ and\ \citenamefont {Trodahl}}]{Holmes-Hewett2019}%
  \BibitemOpen
  \bibfield  {author} {\bibinfo {author} {\bibfnamefont {W.~F.}\ \bibnamefont {Holmes-Hewett}}, \bibinfo {author} {\bibfnamefont {R.~G.}\ \bibnamefont {Buckley}}, \bibinfo {author} {\bibfnamefont {B.~J.}\ \bibnamefont {Ruck}}, \bibinfo {author} {\bibfnamefont {F.}~\bibnamefont {Natali}}, \ and\ \bibinfo {author} {\bibfnamefont {H.~J.}\ \bibnamefont {Trodahl}},\ }\href {\doibase 10.1103/PhysRevB.99.205131} {\bibfield  {journal} {\bibinfo  {journal} {Phys. Rev. B}\ }\textbf {\bibinfo {volume} {99}},\ \bibinfo {pages} {205131} (\bibinfo {year} {2019})}\BibitemShut {NoStop}%
\end{thebibliography}%

\end{document}